\documentclass[times]{TRR}
\usepackage{enumitem}
\usepackage{tabularray}
\usepackage[superscript]{cite}
\newcommand\BibTeX{{\rmfamily B\kern-.05em \textsc{i\kern-.025em b}\kern-.08em
T\kern-.1667em\lower.7ex\hbox{E}\kern-.125emX}}
\def\volumeyear{2024}
\begin{document}
\runninghead{Long et al.}

\title{Remaining useful life prediction of rolling bearings based on refined composite multi-scale attention entropy and dispersion entropy}

\author{Yunchong Long\affilnum{1}, Qinkang Pang\affilnum{1}, Guangjie Zhu\affilnum{1}, Junxian Cheng\affilnum{1} and Xiangshun Li\affilnum{1}}

\affiliation{\affilnum{1}School of Automation, Wuhan University of Technology, Wuhan, China}

\corrauth{Xiangshun Li, School of Automation, Wuhan University of Technology, Wuhan, 430070, China.\\
Email: lixiangshun@whut.edu.cn}

\begin{abstract}
Remaining useful life (RUL) prediction based on vibration signals is crucial for ensuring the safe operation and effective health management of rotating machinery. Existing studies often extract health indicators (HI) from time domain and frequency domain features to analyze complex vibration signals, but these features may not accurately capture the degradation process. In this study, we propose a degradation feature extraction method called Fusion of Multi-Modal Multi-Scale Entropy (FMME), which utilizes multi-modal Refined Composite Multi-scale Attention Entropy (RCMATE) and Fluctuation Dispersion Entropy (RCMFDE), to solve the problem that the existing degradation features cannot accurately reflect the degradation process. Firstly, the Empirical Mode Decomposition (EMD) is employed to decompose the dual-channel vibration signals of bearings into multiple modals. The main modals are then selected for further analysis. The subsequent step involves the extraction of RCMATE and RCMFDE from each modal, followed by wavelet denoising. Next, a novel metric is proposed to evaluate the quality of degradation features. The attention entropy and dispersion entropy of the optimal scales under different modals are fused using Laplacian Eigenmap (LE) to obtain the health indicators. Finally, RUL prediction is performed through the similarity of health indicators between fault samples and bearings to be predicted. Experimental results demonstrate that the proposed method yields favorable outcomes across diverse operating conditions.\\ \\
\textbf{Keywords }\\
Remaining useful life prediction, rolling bearing, health indicator, entropy
\end{abstract}
\maketitle

\section{\textbf{Introduction}}
\noindent The stability and reliability of equipment operation are crucial in various industrial sectors, playing a significant role in ensuring production safety and enhancing economic benefits. Prognostic and Health Management (PHM), also known as Condition-based Maintenance and Predictive Maintenance, offers the advantage of reducing maintenance costs and minimizing unnecessary downtime.\cite{dynamic,ref1} Among the most critical parameters is the Remaining Useful Life (RUL). RUL can be defined as the time remaining from the evaluation point until the machine reaches its end of useful life, with RUL prediction being an estimation of the remaining operating time before machine failure.\cite{ref2} Rolling bearings, as key components in rotating machinery, often experience failures due to harsh working conditions such as heavy loads and high speeds prevalent in industrial applications. Statistics indicate that half of all faults occurring in rotating machinery systems are attributed to rolling bearings.\cite{ref3,ref4,ref5} According to Asea Brown Boveri Ltd, Institue of Electrical and Electronics Engineers and Electric Power Research Institute reports, bearing faults account for 51\% , 41\% and 42\% respectively for induction motor damage caused by rotating machinery.\cite{ref6}  Effective RUL prediction can provide valuable insights and practical references prior to making maintenance decisions, thereby reducing downtime, minimizing economic losses, improving maintenance efficiency and enhancing machine reliability. Consequently, there is an urgent need to develop effective methods for assessing bearings' health status accurately while predicting their Remaining Useful Life.\cite{ref7}

RUL prediction methods for rolling bearings are primarily categorized into model-based approaches and data-driven approaches.\cite{ref8} The model-based methods rely on expert experience and prior knowledge to establish a physical model that accurately reflects the trend of RUL based on the degradation mechanism of rolling bearings. However, due to the increasing complexity of modern mechanical devices, it has become challenging or even impossible to construct such physical models.\cite{ref9} Consequently, promoting model-based methods in industrial applications is difficult. With recent advancements in sensor technology and data transmission technology, data-driven approaches have rapidly emerged.\cite{ref10} This approach mainly utilizes a large amount of sensor-obtained data, which requires minimal reliance on intrinsic degradation mechanisms and prior knowledge of bearings. It offers advantages such as simple implementation, fast response speed and high prediction accuracy, making it suitable for deployment in actual industrial settings. Therefore, both academia and industry have shown great interest in data-driven approaches.\cite{ref8,ref11} Data-driven methodologies can be divided into three steps: data collection, establishment of health indicators and life prediction. Health indicators extract effective features from collected data and transform them into quantifiable indicators that describe equipment degradation processes. Hence, the quality of health indicator construction directly influences the accuracy of RUL prediction.\cite{ref12} 

In order to more accurately quantify the degradation process of equipment performance, numerous scholars have dedicated their efforts to researching methods for constructing robust health indicators.\cite{LSTM,SVM} Generally, health indicators that exhibit good monotonicity and trend properties are considered effective in reflecting the degradation process. The reason is that the degradation of performance is often continuous and irreversible. Yu\cite{ref13} proposed utilizing the local projection algorithm to analyze the manifold structure of time domain and frequency domain indicators, thereby constructing bearing health indicators. By calculating the correlation between 24 time domain, frequency domain and time-frequency domain indicators with lifespan, Guo\cite{ref14} selected 6 sensitive indicators as inputs for an RNN network and employed a constructed RNN-HI model to fuse these input indicators into a health indicator that demonstrates bearing performance degradation. Chen\cite{ref15} extracted 5 bandpass energy values in the frequency domain as degradation features and proposed an RNN with an encoder-decoder framework incorporating attention mechanism for RUL prediction. Wu\cite{extreme} proposed the regularized extreme learning machine based on a staged prediction method and sensitive degradation features. It can be observed from above that degradation features directly extracted from original signals are generally time domain, frequency domain or time-frequency domain statistics. However, most of these features lack obvious monotonicity and trend properties, which hinder accurate reflection of degradation conditions. Therefore, Wang\cite{ref12} directly utilized original vibration signals as inputs for Conv-LSTM neural network models which automatically achieved feature dimension reduction and outputted health indicators by stacking multi-layer networks. Zhu\cite{ref16} put forward a method based on self-attention mechanism and residual hybrid network to construct health indicators for bearings' RUL prediction. Nevertheless,the automatically extracted health indicators exhibit poor interpretability and demonstrate unstable effects in changing working conditions. Additionally, they impose high requirements on data volume.

Entropy is a physical quantity that quantifies the complexity and chaos of a time series. The vibration signal time series of rotating machinery with different degradation levels exhibit different levels of chaos, making entropy useful for extracting degradation features of rotating machinery.\cite{ref17} Entropy can effectively extract deep degradation information from vibration signals and provide stable health indicators with good monotonicity and trend properties. Entropy-based measures such as approximate entropy, sample entropy, fuzzy entropy, and permutation entropy, etc. , have been widely employed in the field of fault diagnosis for rotating machinery.\cite{ref18,ref19,ref20,ref21} Hamed Azami\cite{ref22} proposed a complexity evaluation method for nonlinear time series based on dispersion entropy, which is robust against noise interference and computationally efficient. To further enhance the capability of extracting hidden fault features, Hamed Azami\cite{ref23} subsequently introduced the multi-scale fluctuation dispersion entropy. This method enables more accurate analysis of the chaos of nonlinear time series under different  scale factors compared to aforementioned entropy-based measures. However, it should be noted that inappropriate parameter settings in the calculation algorithm may lead to inaccurate extraction of hidden minor fault features.\cite{ref21} Additionally, existing multi-scale entropy-based measures only analyze low-frequency components while neglecting high-frequency components, thereby overlooking potential fault feature information contained within different frequency segments.\cite{ref24} In light of these limitations, this study conducts relevant research with the following main contributions:

(i) The multi-scale concept is employed to enhance the attention entropy, and a detailed description of RCMATE calculation method and its application in RUL prediction of rolling bearings is provided. This approach enables more accurate and stable extraction of degradation features, while also eliminating dependence on parameter selection.

(ii) We propose a novel method, namely Fusion of Multi-Modal Multi-Scale Entropy(FMME), which combines RCMATE and RCMFDE, for accurately and comprehensively predicting the RUL of rolling bearins.

(iii) The effectiveness of our method is verified under 3 different operating conditions, according to the rolling bearing operating data provided by the IEEE PHM 2012 Challenge. Compared with other methods, the degradation features extracted by our method can reflect the degradation process more accurately. 

The rest of this paper is structured as follows. Section Theoretical backgroung provides a brief introduction to RCMFDE, RCMATE, Laplacian eigenmap and an evaluation metric called MCR for assessing the quality of degradation features. Section The proposed RUL prediction method elaborates on the detailed steps of FMME, the proposed method for predicting RUL of rolling bearings. Section Experimental verification describes the experimental procedure conducted to validate our approach. Finally, section Conclusions summarizes the key findings presented in this study.

\section{\textbf{Theoretical background}}

\subsection{Refined composite multi-scale fluctuation dispersion entropy} 
RCMFDE considers coarse-grained sequences at different time scales, each corresponding to a different starting point of the coarse-grained process. For different displacement sequences, the relative frequencies of fluctuation-based dispersion modes are calculated. Finally, the Shannon entropy of the average occurrence rates of these dispersion modes is defined as RCMFDE.\cite{ref23} It is calculated as follows.\cite{ref21} 

The process begins by dividing the univariate signal  $L: a=\{a_1,a_2,...,a_L\}$ into non-overlapping segments of length  $\tau$ , referred to as the scale factor. Let $k$ denote the number of coarse-grained steps at this scale factor, constructing a coarse-grained time series:
\begin{equation}
\begin{aligned}
    &x_k^{(\tau)}(i)=\frac{1}{\tau}\sum_{c=(i-1)\tau +k}^{i\tau +k-1}a_c , 1\leq i\leq \lfloor \frac{L}{\tau}\rfloor=n,\\ &k=1,2,...,\tau
\end{aligned}
\end{equation}

Next, the standard deviation  $\sigma$ and the mean  $\mu$ of the sequence  $X$ are calculated. Using the normal cumulative distribution function,  $X$ is transformed into  $Y$:
\begin{equation}
    y_k(i)=\frac{1}{\sigma \sqrt{2\pi}}\int_{-\infty}^{x_k(i)}e^{\frac{-(t-\mu )^2}{2\sigma ^2}}dt
\end{equation}

$y_k(i)$ is then linearly mapped to $z_k(i)$ from 1 to $c$ :
\begin{equation}
    z_k^c(i)=round(c\times y_k(i)+0.5)
\end{equation}

Following this, a time delay $d$ and an embedding dimension $(m-1)$ are introduced to form:
\begin{equation}
\begin{aligned}
    &Z_k^{m,c}(j)={z_k^c(j),z_k^c(j+d),...,z_k^c(j+(m-1)d)},\\
    &j=1,2,...,n-(m-1)d
\end{aligned}
\end{equation}

Each sequence $Z_k^{m,c}(j)$ is then transformed into fluctuation-based dispersion modes $\pi_{u_0u_1...u_{m-1}}$ , where  $z_k^c(j+(m-1)d)=u_{m-1}$. For each sequence $Z_k^{m,c}(j)$ , the number of fluctuation-based dispersion modes is $(2c-1)^{m-1}$.

For $Z_k^{m,c}(j)$, the cardinality is denoted by $\#$, and the relative frequency is:
\begin{equation}
\begin{aligned}
    &W(\pi_{a_0a_1...a_{m-1}})=\frac{\#\{j|j\leq n-(m-1)d\}}{n-(m-1)d},\\
    &Z_k^{m,c}(j)hastype \pi_{a_0a_1...a_{m-1}}
\end{aligned}
\end{equation}

Finally, RCMFDE is equal to:
\begin{equation}
-\sum_{\pi=1}^{(2c-1)^{m-1}}\sum_{k=1}^\tau W(\pi_{a_0...a_{m-1}})\times ln(\sum_{k=1}^\tau W(\pi_{a_0...a_{m-1}}))
\end{equation}

The calculation of RCMFDE involves 4 parameters: embedding dimension $m$, class number $c$ ,time delay $d$ , and maximum scale factor $\tau$ . Typically, $d$ is set to 1. Different values of $m$ and $c$ influence the results. Smaller values of $m$ and $c$ result in poorer detection of sudden changes in the signal, while larger values prolong computation time. Therefore, suitable parameter values need to be chosen, with suggested constraints being $(2c-1)^{m-1}\textless \lfloor \frac{L}{\tau_{max}}\rfloor$.

\subsection{Attention entropy}
Attention entropy, proposed by Yang\cite{ref25}, is a novel entropy measure. Traditional entropy measures focus on the frequency distribution of all observations in a time series, whereas attention entropy analyzes the frequency distribution of intervals between key observations. This approach offers faster computation and better performance. Additionally, attention entropy requires no parameter adjustments and is robust to the length of the time series. The specific calculation process is as follows.

The calculation process begins by identifying peak points, which accurately reflect local state boundary changes and are therefore defined as core points.These core points are then assigned according to 4 different strategies: $\{min-min\},\{min-max\},\{max-min\},\{max-max\}$. The intervals between adjacent core points are obtained and denoted as $nn,nx,xn,xx$.

Next, the Shannon entropy of these intervals is calculated using the formula:
\begin{equation}
    H=-\sum_{x=1}^bp(x)log_2 p(x)
\end{equation}
where $p(x)$ is the probability of occurrence of $x$, and $b$ is the number of interval values' types.

Finally, the mean Shannon entropy values obtained from the 4 different strategies are defined as the attention entropy of the time series. The whole calculation process of attention entropy is shown in Figure 1.

\begin{figure}
\centering
\includegraphics[scale=0.35]{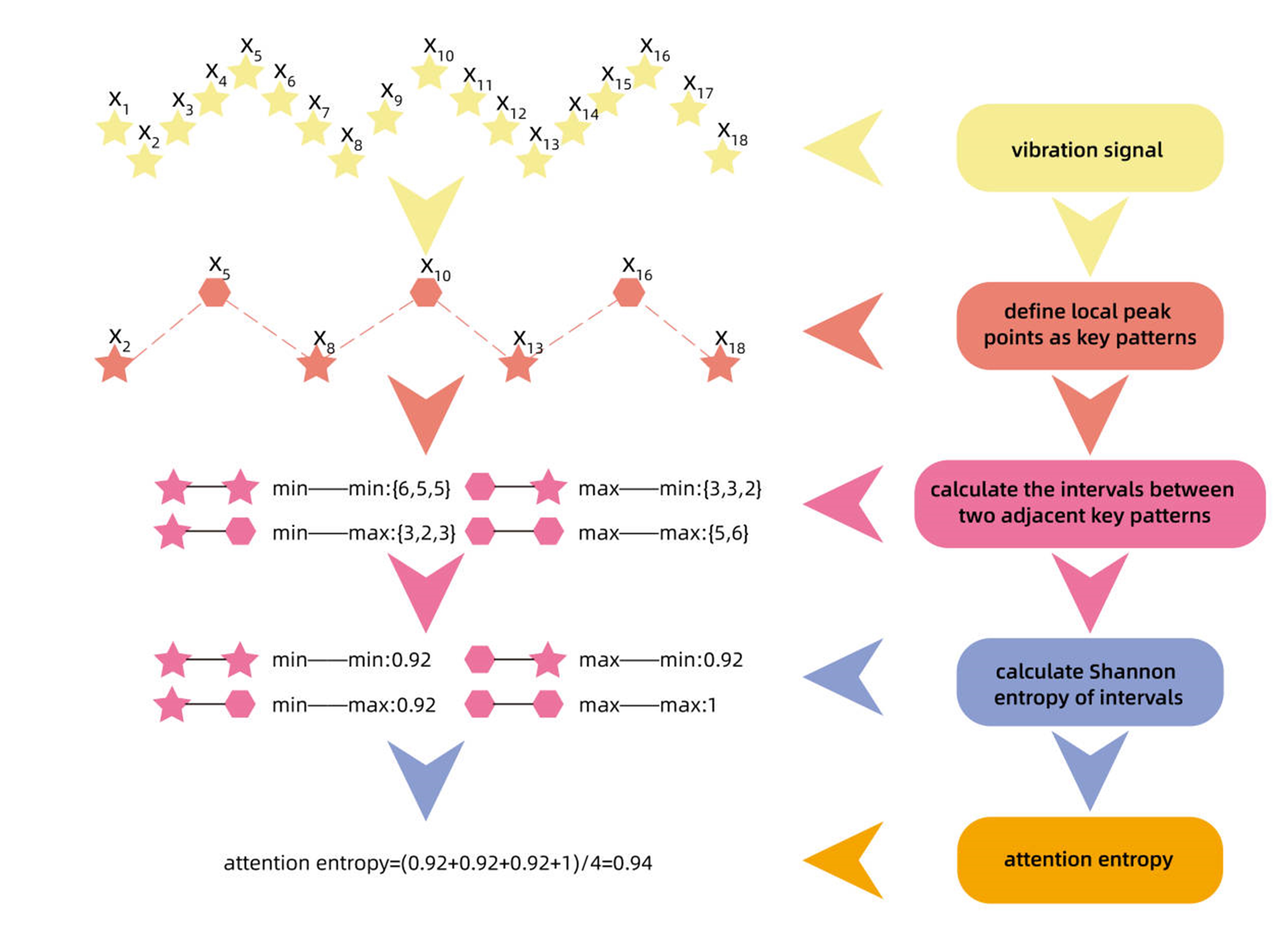}
\caption{Schematic of attention entropy.}
\label{attention entropy}
\end{figure}

\subsection{Refined composite multi-scale attention entropy}
RCMATE considers coarse-grained sequences at different time scales, with each coarse-grained sequence corresponding to a different starting point of the coarse-grained process. For different displacement sequences, the core point intervals are calculated separately under the 4 conditions of $\{min-max\},\{min-min\},\{max-max\},\{max-min\}$. The probability of each type of interval under the 4 conditions is then computed. The average probabilities of the intervals under the 4 conditions across different displacement sequences are calculated. The mean entropy of the 4 average interval probabilities is defined as RCMATE at certain scale. The specific calculation process is as follows.

First, at certain scale factor $j$ , define the number of coarse-grained shifts $w=1,2,...,j$. For each $w$, construct a series of continuous coarse-grained time series:
\begin{equation}
\begin{aligned}
    &X_w^{(j)}(i)=\frac{1}{j}\sum_{c=(i-1)j+w}^{ij+w-1}a_c \\
    &i=1,2,...,\lfloor\frac{L-(w-1)}{j}\rfloor
\end{aligned}
\end{equation}

For different coarse-grained shifts $w$, construct the continuous coarse-grained time series. Then, for each of these sequences, calculate the core point intervals under the 4 conditions of $\{min-max\},\{min-min\},\{max-max\},\{max-min\}$. These core point interval sequences are referred to as $TNX(t),TNN(t),TXX(t)$ and $TXN(t)$. For instance, under the $\{min-max\}$ condition, the core point interval sequence is constructed as:
\begin{equation}
    TNX_w^{(j)}(t)=X_w^{(j)}(t_{max})-X_w^{(j)}(t_{min})
\end{equation}
where $t_{max}$ and $t_{min}$ represent the time indices of the local maxima and minima in the coarse-grained time series, respectively. $t_{max}$ is the smallest index greater than $t_{min}$. 

Next, for each $w$, compute the average of the columns of $TNX_w^{(j)}(t)$ to obtain a row vector. Similarly, calculate the row vectors for the other 3 conditions:
\begin{equation}
\begin{aligned}
    &P1=mean(TNX_w^{(j)}(t)) \\
    &P2=mean(TNN_w^{(j)}(t)) \\
    &P3=mean(TXX_w^{(j)}(t)) \\
    &P4=mean(TXN_w^{(j)}(t)) \\
\end{aligned}
\end{equation}

Then, calculate the entropy of the non-zero values in each of these row vectors, denoted as $E_{nx},E_{nn},E_{xx}$ and $E_{xn}$. Finally, RCMATE at scale $j$ is defined as the mean of these entropy values:
\begin{equation}
    RCMATE_j=\frac{E_{nx}+E_{nn}+E_{xx}+E_{xn}}{4}
\end{equation}

\subsection{Laplacian Eigenmap}
For RUL prediction of rolling bearings, it is essential to collect extensive vibration signal data and extract fault feature information to the greatest extent possible. This inevitably leads to information redundancy. Therefore, it is necessary to perform dimensionality reduction to uncover the essential degradation features contained within the large dataset. Since the vibration data collected by sensors are often non-stationary and nonlinear signals, commonly used linear dimensionality reduction methods, such as Principal Component Analysis (PCA), tend to destroy the topological structure of the vibration data, resulting in the loss of critical information. Consequently, using nonlinear dimensionality reduction algorithms to extract the degradation feature information can more accurately reflect the hidden patterns and essential information within the signals.

Laplacian Eigenmap (LE) is a classical nonlinear manifold learning method. The basic idea of the algorithm is to preserve the local neighborhood information in the data in an average sense during dimensionality reduction. After LE dimensionality reduction, data points that were originally close in the high-dimensional space remain close in the low-dimensional subspace.\cite{ref27} The specific computational steps of this algorithm are not elaborated in this study.

\subsection{Degradation feature evaluation metric MCR}
Before predicting the RUL of rolling bearings, it is necessary to extract degradation features from the vibration signals. Since bearing performance degrades gradually and irreversibly, features with good monotonicity and trend properties are generally considered suitable. Additionally, the robustness of degradation features against external disturbances should also be considered.\cite{ref28,ref29}

\noindent \textbf{Monotonicity coefficient.} \,Due to the fact that degradation features may exhibit an overall increasing or decreasing trend, 2 metrics are considered to measure the monotonicity of the time series $F(t)$:
\begin{equation}
    Mon^{+}(t)=\frac{\sum\limits_{1\leq t_1 \textless t_2 \leq t}\epsilon(F(t_2)-F(t_1))}{\sum\limits_{1\leq t_1 \textless t_2 \leq t}\epsilon(t_2-t_1)}
\end{equation}

\begin{equation}
    Mon^{-}(t)=\frac{\sum\limits_{1\leq t_1 \textless t_2 \leq t}\epsilon(F(t_1)-F(t_2))}{\sum\limits_{1\leq t_1 \textless t_2 \leq t}\epsilon(t_2-t_1)}
\end{equation}
where:\[\epsilon(x)=
\begin{cases}
    1 & x\ge 0 \\
    0 & x\textless 0
\end{cases}
\]

The final monotonicity coefficient:
\begin{equation}
    Mon(t)=max\{Mon^+(t),Mon^-(t)\}
\end{equation}

The closer the value $Mon(t)$ is to 1, the better the monotonicity of the time series.

\noindent \textbf{Correlation coefficient.} \,The correlation coefficient reflects the degree of correlation between the degradation feature series and the actual degradation state of the rolling bearings.\cite{ref30}It is calculated as follows:
\begin{equation}
    Cor=\left| \frac{\sum\limits_{i=1}^n(t_i-\bar{t})(F(t_i)-F(\bar{t}))}{\sqrt{\sum\limits_{i=1}^n(t_i-\bar{t})^2 \cdot \sum\limits_{i=1}^n(F(t_i)-F(\bar{t}))^2}} \right|
\end{equation}

Since the absolute value of the correlation coefficient is taken, it ranges between 0 and 1. The closer the value is to 1, the higher the correlation between the feature and the actual degradation state.

\noindent \textbf{Robustness coefficient.} \,The robustness coefficient is defined as follows:
\begin{equation}
    Rob=\frac{1}{N}\sum\limits_{i=1}^n e^{-\left | \frac{F(t_i)-F(\Tilde{t})}{F(t_i)}\right |}
\end{equation}
where $F(\Tilde{t})$ is the value corresponding to the median index value of the degradation feature time series. The closer $Rob$ is to 1, the smoother the degradation feature changes over time, indicating better robustness against external disturbances.

\noindent \textbf{Comprehensive evaluation metric MCR.} \,The comprehensive evaluation metric $MCR$ integrates monotonicity, trend and robustness of the degradation features, ensuring a more accurate and reliable RUL prediction for rolling bearings. It is calculated as:
\begin{equation}
    MCR=\alpha \cdot Mon+\beta \cdot Cor+\gamma \cdot Rob
\end{equation}
where $\alpha,\beta,\gamma$ are weighting factors that can be adjusted according to practical needs. In this study, the values are set to 0.4, 0.4 and 0.2, respectively.

\begin{figure*}
\centering
\includegraphics[scale=0.23]{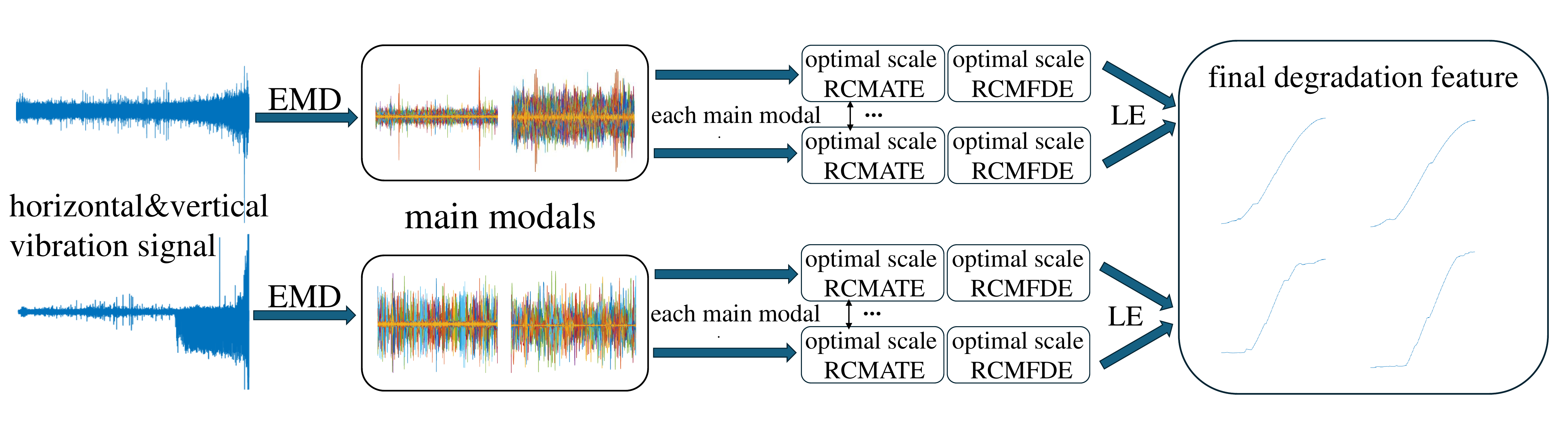}
\caption{The illustration of our FMME method for extracting final degradation features.}
\label{good}
\end{figure*}

\section{\textbf{The proposed RUL prediction method}}
This section details the proposed Fusion of Multi-Modal Multi-Scale Entropy (FMME) method for RUL prediction of rolling bearings, based on EMD and LE, which integrates RCMATE and RCMFDE under multiple modals. The specific process is as follows.

To begin with, accelerometers are used to collect vibration signals of rolling bearings in horizontal and vertical directions. The accelerometers  collect $x$ data points at intervals of time $T$, and this is repeated $n$ times, taking a total duration of $nT$.

Next, the original vibration data is decomposed into multiple modals using EMD, and the main modals are selected for further analysis. For simplicity, the first 6 modals are directly selected in this study. By processing the vibration signals' multiple main modals, i.e. , the major components in different frequency bands, it is possible to fully exploit the hidden fault information in each frequency band while discarding some noise components.

For each modal of the vibration signal at each data acquisition time, the RCMATE and RCMFDE are calculated for each sequence of $x$ data points over 20 scales. This process is repeated $n$ times to obtain sequences of RCMATE and RCMFDE over time for 20 scales. If the number of scales selected is too small, the time series information cannot be fully utilized. Conversely, if the number of scales is too large, the computation time becomes excessively long. Therefore, this study selects 20 scales as a compromise.

Subsequently, wavelet denoising\cite{wavelet} and exponential sliding window smoothing are applied to these entropy sequences, and the comprehensive evaluation metric MCR for each sequence corresponding to different scales is then calculated. Since the degradation feature time series obtained directly contain significant noise and fluctuations, smoothing and denoising are necessary. In this study, the sliding window size is set to 50, with 3 sliding iterations empirically. After denoising, the degradation feature time series generally exhibit good monotonicity, trend and robustness.

In each main modal, the RCMATE and RCMFDE at the scale with the highest MCR are selected. Using LE, the optimal scale horizontal RCMATE, optimal scale horizontal RCMFDE, optimal scale vertical RCMATE and optimal scale vertical RCMFDE from the 6 main modals are fused to obtain 4 final fused degradation feature sequences, which is shown in Figure 2. The images of the final degradation feature sequences after LE dimensionality reduction should exhibit good symmetry and monotonicity about $y=0$ , corresponding to the continuous and irreversible nature of rolling bearing performance degradation. Sequences with poor symmetry or monotonicity are not used as final degradation features, as this results in poor RUL prediction accuracy. The comparison plots of good and poor features are shown in Figure 3.

\begin{figure}
\centering
\includegraphics[scale=0.35]{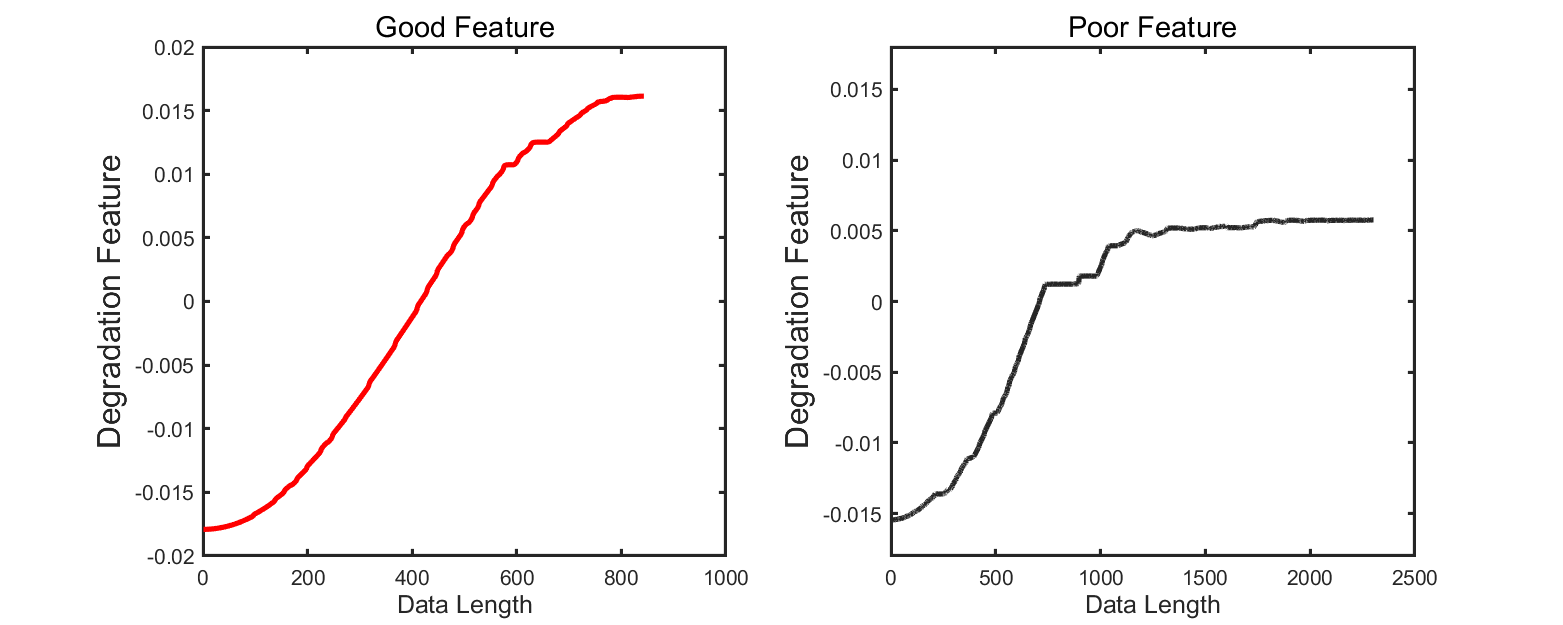}
\caption{Comparison of good and poor features.}
\label{good}
\end{figure}

For the vibration data of the fault samples, the RCMATE and RCMFDE at the scale with the highest MCR in the horizontal and vertical directions are used as the 4 final degradation feature sequences, which is the same as the above steps. The time points where the final degradation feature sequences' values show abrupt changes are identified as the times of fault occurrence. During normal operation, the vibration signal amplitude is relatively stable, with minor changes. But it increases sharply in a short time before a fault occurs. The final degradation feature time sequences extracted using our method show an obvious monotonic trend during the normal operation phase of the rolling bearing, reflecting its continuous degradation process. Before a fault occurs, the trend reverses and changes sharply. This turning point is regarded as the bearing failure time, as shown in Figure 4, which often occurs slightly earlier than the sharp increase in vibration signal amplitude, making it more conducive to timely fault prediction and maintenance, aligning with the conservative characteristics of engineering.

\begin{figure}
\centering
\includegraphics[scale=0.32]{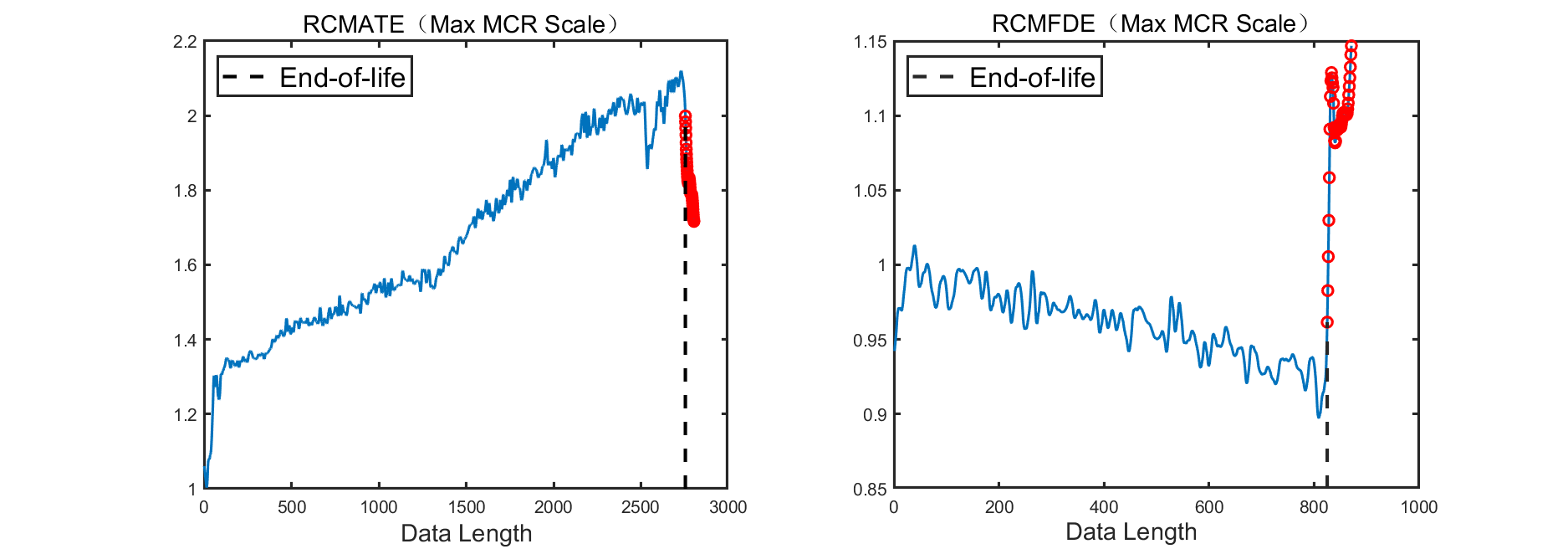}
\caption{Schematic of bearing failure points.}
\label{end}
\end{figure}

The average of the 4 failure times is defined as the failure time $T_{failure}$ of the bearing. A health indicator label is attached to each time point $t$ corresponding to the final degradation feature sequences, calculated as follows:
\begin{equation}
    label(t)=\frac{T_{failure}-t}{T_{failure}}
\end{equation}
Evidently, as the bearing operates, this value decreases from 1 to 0, reflecting the degradation process of bearing performance.

For each final degradation feature sequence of the bearing to be predicted, a coefficient $K_i$ is multiplied so that the initial value of each final degradation feature sequence matches the initial value of the corresponding final degradation feature sequence of the fault sample. This method aims to compare the final degradation features of the predicted bearing with those of the fault sample on the same scale, using the similarity between the two to estimate the RUL of the predicted bearing. 

For the current value of each final degradation feature sequence of the predicted bearing, the corresponding closest value in the fault sample is found. The average of the successfully matched time points is taken as the current health indicator value $present\_label$ . A straight line is then fitted through the two points $(1,1)$ and $(present\_time,present\_label)$, with the intersection of the line and $y=0$ being the estimated failure time $t_{fail}$ of the bearing. The RUL is then calculated as:
\begin{equation}
    RUL=t_{fail}-present\_time
\end{equation}

\section{\textbf{Experimental verification}}
In this section, the proposed RUL prediction method is applied to the rolling bearing vibration signals from IEEE PHM 2012 Challenge. Our method is tested under 3 different operating conditions and compared with several typical methods to evaluate its effectiveness and superiority.

\begin{figure*}
\centering
\includegraphics[scale=0.4]{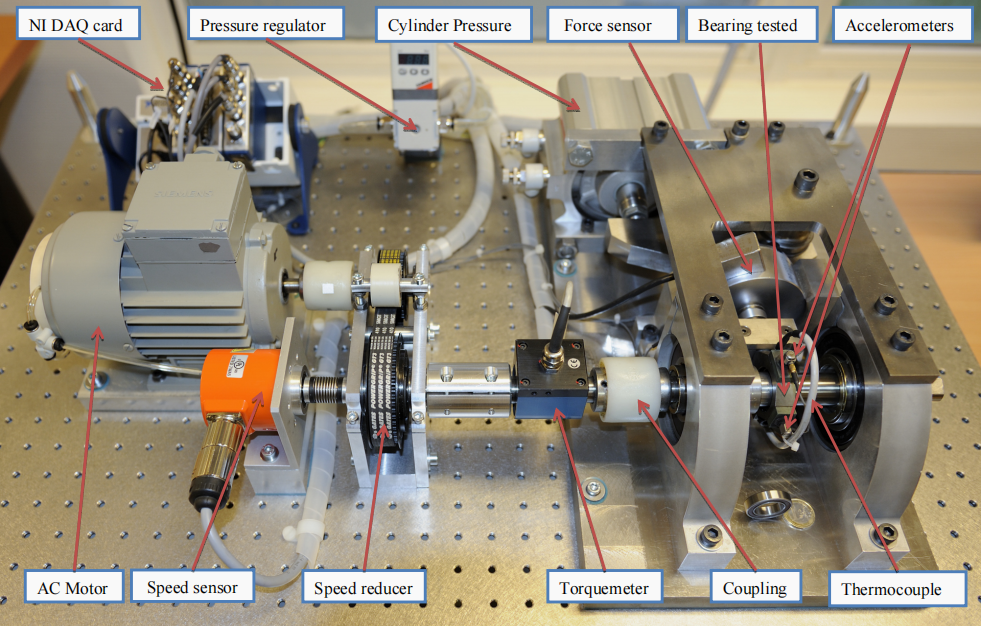}
\caption{IEEE PHM 2012 Challenge bearing accelerated degradation experimental platform.}
\label{platform}
\end{figure*}

\subsection{Experimental platform}
The experimental platform, as shown in Figure 5, was designed and implemented by the AS2M department of the FEMTO-ST Institute.\cite{IEEE} It provides real accelerated degradation experimental data from rolling bearings operating until failure. The platform consists of 3 main parts: the rotating section, the degradation induction section and the measurement section. The rotating section primarily includes an asynchronous motor and a gearbox, with the motor having a power of 250W and a maximum rotational speed of 2830rpm, which transmits rotational motion to drive the bearings. The degradation induction section, the core of the experimental platform, employs a pneumatic jack to apply dynamic loads to the bearings, accelerating their degradation. The measurement section mainly includes a resistance temperature detector and 2 accelerometers, which are placed on the outer ring of the bearing to measure vibration signals in the horizontal and vertical directions. The sampling frequency is 25.6kHz, with data collected every 10 seconds for a duration of 0.1 seconds per sample, resulting in 2560 data points per sample.

\subsection{Dataset description}
The dataset used in this study contains bearing vibration data under 3 different operating conditions as shown in Table 1. The competition dataset consists of 2 parts: the learning set, which contains fault sample data, and the test set, which includes truncated data.

\begin{table}
\centering
\caption{Bearing working condition.}
\scalebox{0.78}{
\begin{tblr}{
  cells = {c},
  hline{1-2,5} = {-}{},
}
Datasets       & Condition 1                & Condition 2                                          & Condition 3   \\
Load and speed & 4000N 1800rpm              & 4200N 1650rpm                                        & 5000N 1500rpm \\
Learning set   & {Bearing1\_1\\Bearing1\_2} & {Bearing2\_1\\Bearing2\_2}                           & Bearing3\_1   \\
Test set       & {Bearing1\_3\\Bearing1\_7} & {Bearing2\_3\\Bearing2\_4\\Bearing2\_5\\Bearing2\_7} & Bearing3\_3   
\end{tblr}
}
\end{table}
Wang, the runner-up in the industrial group of the IEEE PHM 2012 Challenge, pointed out that the competition defined the failure moment of the bearing as when the vibration signal amplitude exceeds 20g. However, at this point, the bearing's amplitude is so large that it threatens the safety of the entire experimental platform. Moreover, using this definition makes the RUL prediction problem essentially about predicting the first time the original vibration signal amplitude crosses a certain threshold, which is unreasonable.\cite{ref31} Bearing amplitude is usually very stable until it approaches failure, where it then changes drastically, making this process poorly predictable. Therefore, this study adopts a failure definition that is more conducive to system safety and prediction accuracy. Specifically, the failure moment is defined as the average time when RCMATE and RCMFDE exhibit a sharp abnormal change at the maximum MCR scale. This moment is beneficial for system safety and more consistent with the bearing degradation pattern. The failure moments of all bearings and the truncation moments of the test set bearings are shown in Table 2.

\begin{table}
\centering
\caption{The failure moments of all bearings and the truncation moments of the test set bearings.}
\scalebox{0.6}{
\begin{tblr}{
  cells = {c},
  cell{1}{1} = {r=2}{},
  cell{1}{6} = {r=2}{},
  cell{1}{7} = {r=2}{},
  cell{2}{2} = {c=4}{},
  hline{1,3,15} = {-}{},
}
Bearing & X RCMATE  & Y RCMATE & X RCMFDE & Y RCMFDE & Average & {Truncation\\time(10s)} \\
        & Life(10s) &          &          &          &         &                         \\
1\_1    & 2754      & 2730     & 2749     & 2725     & 2739    & --                      \\
1\_2    & 862       & 833      & 852      & 825      & 843     & --                      \\
1\_3    & 2249      & 2259     & 2304     & 2295     & 2277    & 1802                    \\
1\_7    & 2102      & 2027     & 2201     & 2028     & 2090    & 1502                    \\
2\_1    & 886       & 872      & 873      & 872      & 876     & --                      \\
2\_2    & 748       & 745      & 747      & 745      & 746     & --                      \\
2\_3    & 1930      & 1935     & 1934     & 1938     & 1934    & 1201                    \\
2\_4    & 735       & 732      & 731      & 730      & 734     & 611                     \\
2\_5    & 2310      & 2310     & 2263     & 2258     & 2285    & 2001                    \\
2\_7    & 214       & 216      & 192      & 214      & 209     & 171                     \\
3\_1    & 488       & 487      & 478      & 480      & 483     & --                      \\
3\_3    & 417       & 414      & 433      & 419      & 421     & 351                     
\end{tblr}
}
\end{table}
\subsection{Scoring of results}
The competition scores based on the following criteria: Let the actual RUL be denoted as $ActRUL$ and the estimated RUL be denoted as $RUL$. For each bearing to be predicted, the error percentage is calculated as follows:
\begin{equation}
    \%Er=100\times\frac{ActRUL-RUL}{ActRUL}
\end{equation}
If the error is positive, it indicates that the RUL prediction is early. If the error is negative, it indicates that the RUL prediction is late. The scoring standards for the two cases are different. The score for each bearing is calculated as follows:
\begin{equation}
A_i=
\begin{cases}
    e^{-ln(0.5)\cdot (\frac{Er_i}{5})}, & Er_i\leq 0 \\
    e^{ln(0.5)\cdot (\frac{Er_i}{20})}, & Er_{i}  \,\textgreater \, 0
\end{cases}
\end{equation}
The final score is the average score of all bearings:
\begin{equation} 
Score=\frac{1}{n}\sum_{i=1}^{n}A_i
\end{equation}
The graph corresponding to Equation (21) is shown in Figure 6.
\begin{figure}
\centering
\includegraphics[scale=0.45]{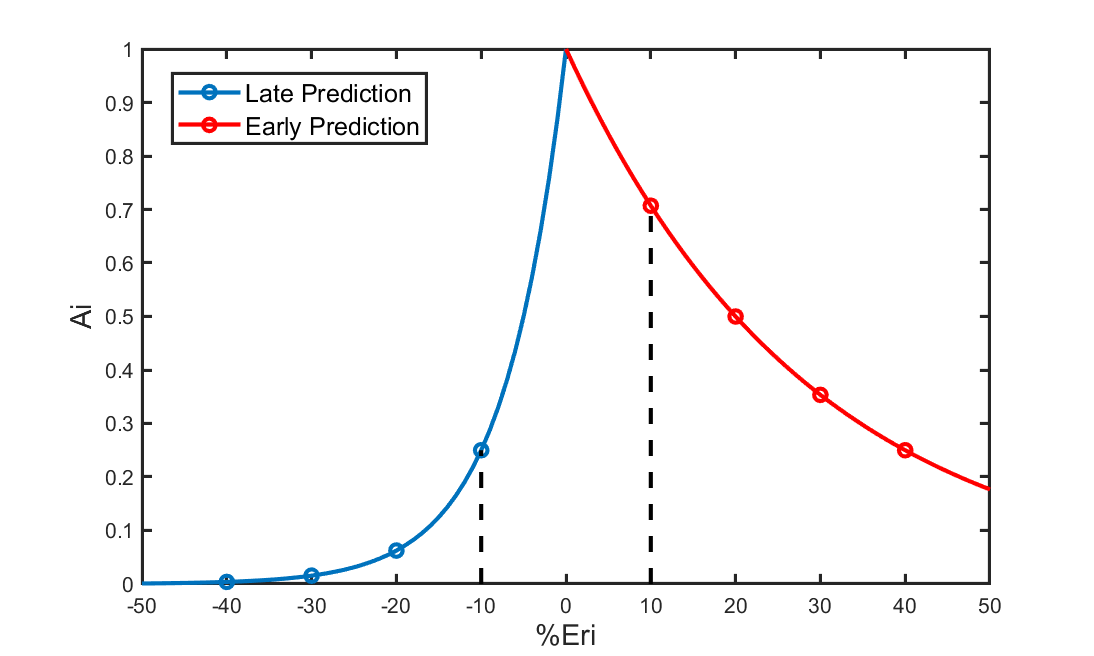}
\caption{Schematic diagram of the scoring function.}
\label{score}
\end{figure}
\subsection{Degradation feature evaluation}
In this section, we take the optimal scale RCMATE and RCMFDE of the fault sample bearing within the first 20 scales as examples to demonstrate their graphs and MCR values. These are compared with commonly used time-domain features in bearing vibration signals, all of which have undergone wavelet denoising and exponential sliding window smoothing, to prove the effectiveness and superiority of the entropy measures used in this study in reflecting the degradation state before bearing failure.
\begin{table}
\centering
\caption{MCR values of different features.}
\scalebox{0.62}{
\begin{tblr}{
  cells = {c},
  hline{1-2,8} = {-}{},
}
Bearing & RCMATE  & RCMFDE & Mean    & Peak    & Kurtosis & Std     & ImpulseFactor \\
1\_1    & 0.9562  & 0.9443 & 0.4022  & 0.8181  & 0.892    & 0.8106  & 0.9369        \\
1\_2    & 0.9499  & 0.9423 & 0.4215  & 0.756   & 0.7496   & 0.7555  & 0.8191        \\
2\_1    & 0.8946  & 0.8047 & 0.4358  & 0.9079  & 0.7655   & 0.702   & 0.7322        \\
2\_2    & 0.8923  & 0.9082 & 0.6571  & 0.8832  & 0.8917   & 0.8715  & 0.8748        \\
3\_1    & 0.8819  & 0.889  & 0.6972  & 0.8704  & 0.7996   & 0.854   & 0.8102        \\
Average & 0.9150 & 0.8977 & 0.5228 & 0.8471 & 0.8197  & 0.7987 & 0.8346       \\
        &         &        &         &         &          &         &               
\end{tblr}
}
\end{table}
As shown in Table 3, the average MCR values of the optimal scale RCMATE and RCMFDE for the 5 bearings are significantly higher than the average MCR values of the 5 commonly used time-domain features: mean, peak, kurtosis, standard deviation and impulse factor. This indicates that RCMATE and RCMFDE, when used as degradation features, have better monotonicity, trend and robustness, making them more effective in reflecting the degradation state before bearing failure. Additionally, the MCR values of RCMATE and RCMFDE are relatively stable under different operating conditions and across different bearings, with maximum fluctuations of 0.0743 and 0.1396, respectively. In contrast, the maximum fluctuations of the MCR values for the 5 time-domain features are 0.295, 0.1519, 0.1424, 0.1695 and 0.2047, respectively, demonstrating that RCMATE and RCMFDE have stronger generalizability.
\begin{figure}
\centering
\includegraphics[scale=0.35]{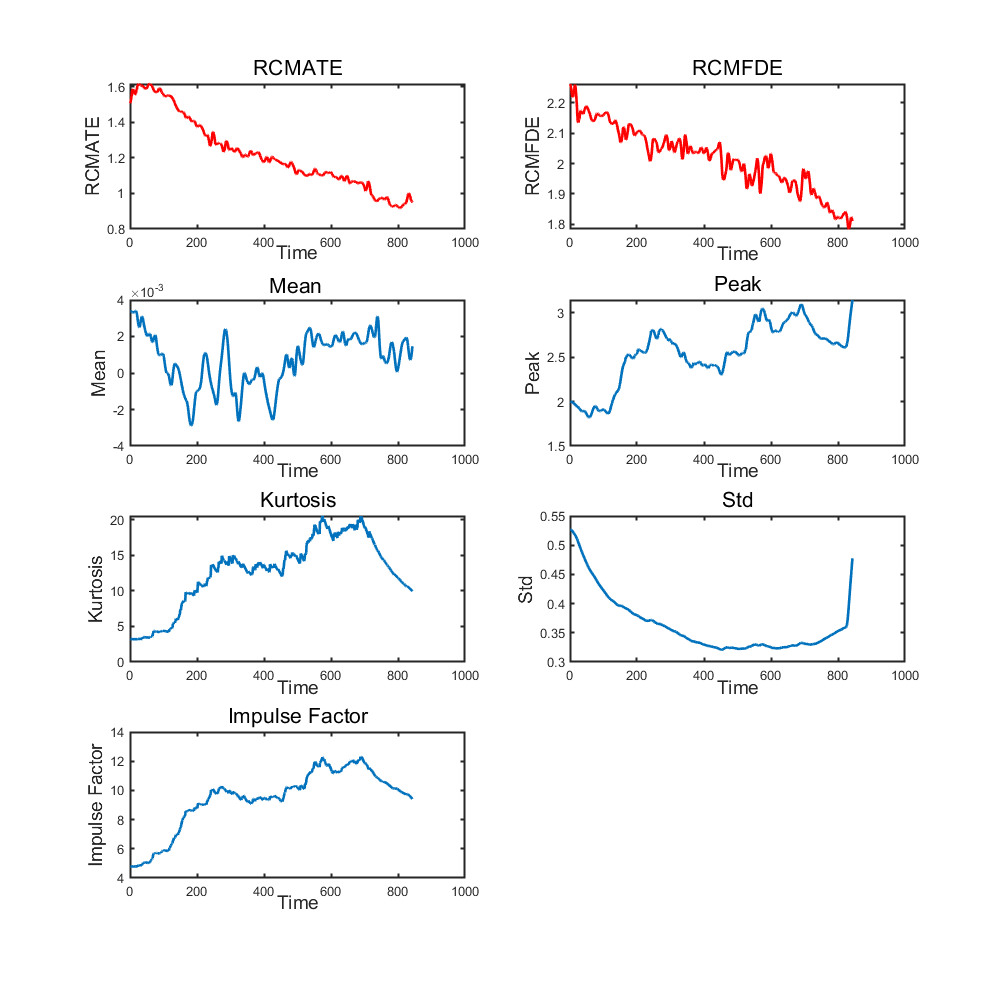}
\caption{Schematic diagram of the variation of different features over time for bearing 1\_2.}
\label{feature}
\end{figure}
As shown in Figure 7, the trends of RCMATE and RCMFDE at the optimal scale are stable with little fluctuation before bearing failure, accurately reflecting the degradation process of bearing. In contrast, the mean, peak, kurtosis and impulse factor exhibit large fluctuations, with poor monotonicity, trend and robustness. The standard deviation shows very small fluctuations, with a stable decrease in the early stage but a plateau in the mid-to-late stages, failing to accurately reflect the actual degradation state of the bearing.

\subsection{RUL prediction}
According to the method described above, the 4 final degradation features of the fault samples are calculated after LE, along with the health status labels, as shown in Figure 8.
\begin{figure}
\centering
\includegraphics[scale=0.34]{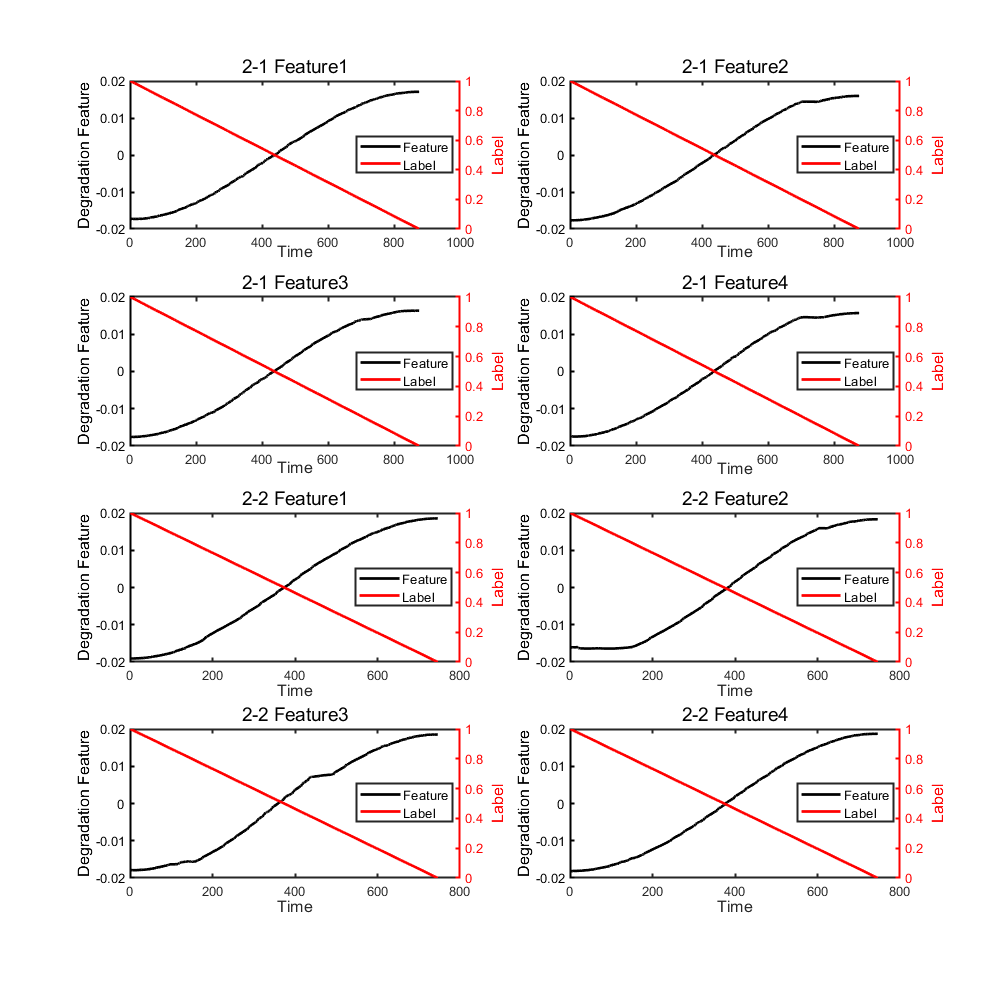}
\caption{The 4 final degradation features and corresponding health labels obtained after LE for bearing 2\_1 and 2\_2.}
\label{feature2}
\end{figure}
From Figure 8, it can be seen that most of the final degradation features obtained after LE of the fault samples exhibit good monotonicity and symmetry, with similar values and shapes. This indicates that our method can extract the deep degradation patterns of bearings, revealing the essential and common degradation characteristics of different bearings.

Next, the final degradation features of the truncated data of the bearings to be predicted are calculated after LE, discarding the few features with poor symmetry. As shown in Figure 9, a comparison of good and poor degradation features after LE for bearing 2\_5 is presented.
\begin{figure}
\centering
\includegraphics[scale=0.38]{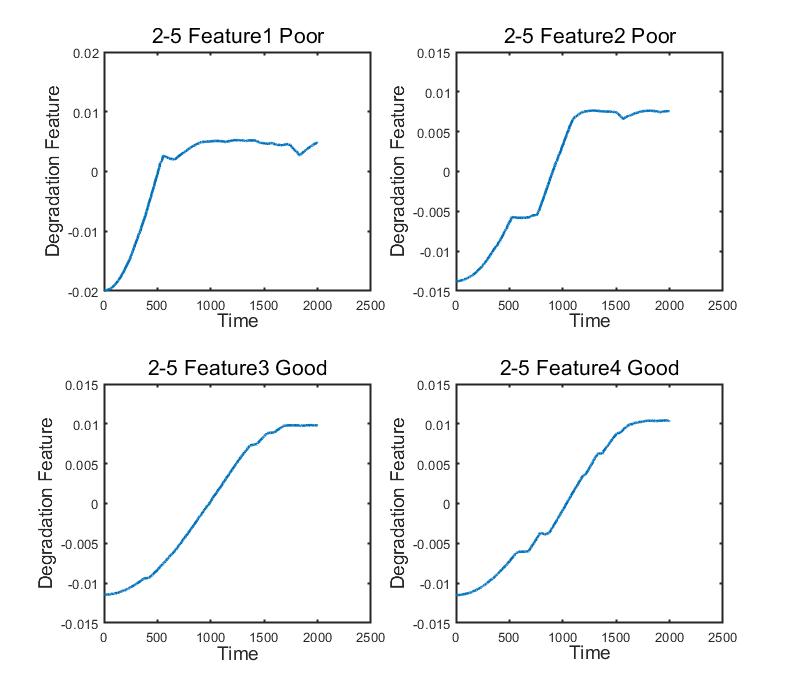}
\caption{A comparison of good and poor degradation features after LE for bearing 2\_5.}
\label{good2}
\end{figure}
In operating condition 1 and 2, there are 2 fault samples each, and in operating condition 3, there is 1 fault sample. For the bearings to be predicted, each final degradation feature obtained after LE needs to be multiplied by a coefficient when compared with the fault samples, so that the initial value of each final degradation feature matches the initial value of the corresponding feature of the fault sample. Since the final degradation features obtained through our method have good symmetry and similar shapes across different bearings, this method allows for the comparison of the final degradation features of different bearings on the same scale. Then, each feature value of the bearing to be predicted at the truncation moment is searched in the corresponding feature of the fault samples. If the search is successful, the label value corresponding to the feature value in the fault sample is taken as the estimated health indicator of the bearing at the truncation moment, and all obtained estimates are averaged. Searches that fail are marked as $\times$, and features that do not participate in the RUL prediction are marked as $--$. The estimated health indicator values are shown in Table 4.
\begin{table}
\centering
\caption{Estimated health indicators of test set bearings at truncation moments obtained from fault samples.}
\scalebox{0.65}{
\begin{tblr}{
  cells = {c},
  cell{1}{2} = {c=2}{},
  cell{1}{4} = {c=2}{},
  cell{1}{6} = {c=2}{},
  cell{1}{8} = {c=2}{},
  cell{8}{2} = {c=2}{},
  cell{8}{4} = {c=2}{},
  cell{8}{6} = {c=2}{},
  cell{8}{8} = {c=2}{},
  hline{1-2,9} = {-}{},
}
Target & Feature1 &    & Feature2 &        & Feature3 &        & Feature4 &       & Average \\
1\_3   & 0.1402   & ×  & ×        & ×      & 0.2663   & --     & ×        & --    & 0.2033  \\
1\_7   & 0.1041   & ×  & 0.0796   & 0.0855 & 0.0807   & --     & ×        & --    & 0.0875  \\
2\_3   & ×        & ×  & ×        & 0.2282 & 0.2149   & 0.2175 & 0.2731   & 0.266 & 0.2399  \\
2\_4   & 0.048    & ×  & ×        & 0.2054 & ×        & ×      & ×        & ×     & 0.1267  \\
2\_5   & --       & -- & --       & --     & 0.1303   & 0.1772 & 0.04     & 0.162 & 0.1274  \\
2\_7   & 0.0571   & ×  & ×        & 0.192  & ×        & 0.0765 & ×        & 0.156 & 0.1204  \\
3\_3   & ×        &    & ×        &        & ×        &        & 0.2075   &       & 0.2075  
\end{tblr}
}
\end{table}
Next, a straight line is fitted through points $(1,1)$ and $(present\_time,present\_label)$, with the intersection of this line and $y=0$ representing the predicted failure time. The predicted failure time minus the truncation time gives the RUL prediction value. The scores are calculated as shown in Table 5.
\begin{table}
\centering
\caption{RUL prediction results, errors and scores.}
\scalebox{0.90}{
\begin{tblr}{
  cells = {c},
  hline{1-2,9} = {-}{},
}
Target & Pre\_RUL (10s) & Act\_RUL (10s) & Err      & Score  \\
1\_3   & 460            & 475            & 3.16\%   & 0.8963 \\
1\_7   & 144            & 588            & 75.51\%  & 0.073  \\
2\_3   & 354            & 733            & 51.70\%  & 0.1667 \\
2\_4   & 89             & 123            & 27.64\%  & 0.3837 \\
2\_5   & 293            & 284            & -3.17\%  & 0.6444 \\
2\_7   & 24             & 38             & 36.84\%  & 0.2789 \\
3\_3   & 92             & 70             & -31.43\% & 0.0128 
\end{tblr}
}
\end{table}
\subsection{Comparison and analysis}
\begin{table}
\centering
\caption{Comparison of scores of different methods.}
\scalebox{0.85}{
\begin{tblr}{
  cells = {c},
  hline{1-2,9} = {-}{},
}
Target  & Proposed & Time-Frequency & RNN-HI & Conv-LSTM \\
1\_3    & 0.8963   & 0.2774         & 0.2231 & 0.3112    \\
1\_7    & 0.073    & 0.7579         & 0.5391 & 0.6329    \\
2\_3    & 0.1667   & 0.1088         & 0.2694 & 0.0106    \\
2\_4    & 0.3837   & 0.7071         & 0.0677 & 0.0011    \\
2\_5    & 0.6444   & 0              & 0.1519 & 0.004     \\
3\_3    & 0.0128   & 0.0442         & 0.8809 & 0.7759    \\
Average & 0.3508   & 0.2701         & 0.3046 & 0.2543    
\end{tblr}
}
\end{table}
The RUL prediction scores obtained using our method are compared with 3 other representative methods, as shown in Table 6. The average score of our method is significantly higher than that of the other 3 methods. The first method\cite{ref32}, the academic group champion of the IEEE PHM 2012 Challenge, uses a time-frequency domain analysis method, which is quite representative. Time-frequency domain analysis performs well on bearings with good monotonicity and trend properties in certain time-domain and time-frequency domain features. However, for many bearings, it is difficult to find satisfactory features from classical time-domain and time-frequency domain features, which makes this method quite limited. The second method\cite{ref14} evaluates classical time-frequency features using monotonicity , correlation and other indicators, selecting features sensitive to the degradation process as inputs for the RNN. Although this method introduces neural networks, it still does not solve the limitation that classical time-frequency features cannot accurately reflect the degradation process. The third method\cite{ref12} directly uses the vibration signals as inputs for the Conv-LSTM, automatically achieving feature dimensionality reduction and health indicator output without prior knowledge. However, the vibration signals do not intuitively reflect the degradation patterns of bearings and contain much noise. The redundant information can reduce overall prediction accuracy. In contrast, the degradation features used in our proposed method can accurately reflect the degradation process of bearings, reveal the deep fault information and common degradation patterns of bearings, and thus achieve better RUL prediction results. 

\section{\textbf{Conclusion}}
Extracting degradation features from the original vibration signals of rolling bearings that can accurately reflect the degradation process is crucial for RUL prediction. To overcome the limitations of traditional feature extraction methods in capturing deep degradation information from vibration signals, this study proposes a method called Fusion of Multi-Modal Multi-Scale Entropy (FMME) for rolling bearing degradation feature extraction and RUL prediction. RCMATE and RCMFDE are capable of more accurately reflecting the degradation process of bearings and have strong generalizability. By using EMD and LE to fuse RCMATE and RCMFDE across multiple modals, the method overcomes the limitations of multi-scale entropy in reflecting high-frequency information of vibration signals while extracting important degradation information and avoiding information redundancy. The final degradation features obtained through LE show high similarity across different bearings, enabling the extraction of the essential degradation patterns of bearings. By comparing the final degradation features of fault samples with those of the bearings to be predicted, the RUL of the bearings can be obtained. Extensive experiments conducted on multiple bearings under 3 different operating conditions demonstrate the superiority of the proposed method over traditional feature extraction methods and its effectiveness in RUL prediction.

Additionally, our research utilizes the similarity between the degradation features of fault samples and the bearings to be predicted for RUL prediction. Future work will delve into transfer learning to fully explore the similarities between different bearings for more accurate RUL prediction. However, in reality, rotating machinery often consists of multiple coupled components. Therefore, extending RUL prediction from individual components to the entire system will also be a focus of future work.

\section{\textbf{Declaration of Conflicting Interests}}
The author(s) declared no potential conflicts of interest with respect to the research, authorship, and/or publication of this article.

\section{\textbf{Funding}}
The author(s) received no financial support for the research, authorship, and/or publication of this article.

\bibliographystyle{TRR}
\renewcommand\refname{Reference}
\bibliography{reference}

\end{document}